\documentclass{article}
\usepackage{moreverb}
\usepackage{url}
\usepackage{grffile}
\usepackage{lipsum}
\usepackage[UKenglish]{isodate}
\usepackage{natbib}
 \usepackage{geometry}
 \usepackage{amsmath}
\usepackage{amssymb}
\usepackage{graphicx} % support the \includegraphics command and options

\usepackage{fancyhdr} % This should be set AFTER setting up the page geometry
\newcommand\BibTeX{{\rmfamily B\kern-.05em \textsc{i\kern-.025em b}\kern-.08em
T\kern-.1667em\lower.7ex\hbox{E}\kern-.125emX}}
\newcommand\MiKTeX{{\rmfamily M\kern-.05em \textsc{i\kern-.025em K}\kern-.08em
T\kern-.1667em\lower.7ex\hbox{E}\kern-.125emX}}
\newcommand\PracTeX{{\rmfamily P\kern-.05em \textsc{r\kern-.025em a\kern-.025em
c}\kern-.08em
T\kern-.1667em\lower.7ex\hbox{E}\kern-.125emX}}
\newcommand{\bG}{\mbox{$\boldsymbol{G}$}}
\newcommand{\bI}{\mbox{$\boldsymbol{I}$}}
\newcommand{\bK}{\mbox{$\boldsymbol{K}$}}
\newcommand{\bc}{\mbox{$\boldsymbol{c}$}}
\newcommand{\boldf}{\mbox{$\boldsymbol{f}$}}
\newcommand{\bh}{\mbox{$\boldsymbol{h}$}}
\newcommand{\bs}{\mbox{$\boldsymbol{s}$}}
\newcommand{\bu}{\mbox{$\boldsymbol{u}$}}
\newcommand{\bw}{\mbox{$\boldsymbol{w}$}}
\newcommand{\bx}{\mbox{$\boldsymbol{x}$}}
\newcommand{\by}{\mbox{$\boldsymbol{y}$}}
\newcommand{\bz}{\mbox{$\boldsymbol{z}$}}
\newcommand{\bX}{\mbox{$\boldsymbol{X}$}}
\newcommand{\bfalpha}{\boldsymbol{\alpha}}
\newcommand{\bbeta}{\boldsymbol{\beta}}
\newcommand{\bfvarepsilon}{\boldsymbol{\varepsilon}}
\newcommand{\bfdelta}{\boldsymbol{\delta}}
\newcommand{\thetab}{\boldsymbol{\theta}}

\newcommand{\bfmu}{\boldsymbol{\mu}}
\newcommand{\bSigma}{{\boldsymbol\Sigma}}
\newcommand{\bxi}{{\boldsymbol\xi}}
\newcommand{\bzero}{{\textbf 0}}

\newcommand{\corr}{\textrm{corr}}
\newcommand{\var}{\textrm{var}}
\newcommand{\cov}{\textrm{cov}}
\newcommand{\RR}{\mathbb{R}}
\begin{document}
\cleanlookdateon
% Specifying a running head, title, author(s), and affiliation
% in the manner required by the Journal.
% Note that the first argument of "\runningheads" specifies the
% "brief" running title (which will appear on the even numbered
% pages) and the second argument specifies the format in which
% the authors' names will appear on the odd numbered pages.
% Do *not* use "\corrauth" if there is only one author.
% *Do* use "\addressnum{1}" even if there is only one author!
%\runningheads{A few statistical principles for data science}{N~CRESSIE}
\title{A few statistical principles for data science}
\author{Noel Cressie\\National Institute for Applied Statistics Australia\\University of Wollongong\\Wollongong NSW 2522, Australia\\email: ncressie@uow.edu.au}
%\affiliation{University of Wollongong
%}
% Specifying address(es) in the manner required by the Journal.
%\address{
%\addressnum{1} \hspace*{-2ex} National Institute for Applied Statistics Research Australia (NIASRA)\\University of Wollongong\\Wollongong NSW 2522, Australia\\
%\hspace*{1ex} Email: \texttt{ncressie@uow.edu.au}
%}
% Note that the Journal requires that a paper must begin with a
% "Summary" not an "Abstract".  This is automatically taken care
% of by the anzsauth document style.  So even though the following
% says "\begin{abstract}" the heading "Summary" will appear in
% the processed version.

\maketitle

\begin{abstract}
In any other circumstance, it might make sense to define the extent of the terrain (Data Science) first, and then locate and describe the landmarks (Principles).
But this data revolution we are experiencing defies a cadastral survey.
Areas are continually being annexed into Data Science.
For example, biometrics was traditionally statistics for agriculture in all its forms but now, in Data Science, it means the study of characteristics that can be used to identify an individual.
Examples of non-intrusive measurements include height, weight, fingerprints, retina scan, voice, photograph/video (facial landmarks and facial expressions), and gait.
A multivariate analysis of such data would be a complex project for a statistician, but a software engineer might appear to have no trouble with it at all.
In any applied-statistics project, the statistician worries about uncertainty and quantifies it by modelling data as realisations generated from a probability space.
Another approach to uncertainty quantification is to find similar data sets, and then use the variability of results between these data sets to capture the uncertainty.
Both approaches allow `error bars' to be put on estimates obtained from the original data set, although the interpretations are different.
A third approach, that concentrates on giving a single answer and gives up on uncertainty quantification, could be considered as Data Engineering, although it has staked a claim in the Data Science terrain.
This article presents a few (actually nine) statistical principles for data scientists that have helped me, and continue to help me, when I work on complex interdisciplinary projects.


\end{abstract}

% Note that "keywords" should not include words and phrases
% that form part of the title of the paper.
% Also keywords (with the exception of proper names and
% certain abbreviations that conventionally appear all in
% capital letters) should *not* be capitalised.

% This shows how acknowledgements should appear in the Journal.
% Note that if you are acknowledging financial support, grant
% numbers should *not* usually be specified.

\section{Does your analysis have `error bars'?}

Science and Engineering have long been held separate by our universities and learned academies.
Broadly speaking, Science is concerned with `why,' and Engineering is concerned with `how.'
Data Science seems to include both~\citep{cressie2020}, and sometimes it may be hard to find the science in a Data Science solution.
For example, an idea is born on how a particular data set could be analysed, perhaps with only a vague idea of the question being answered.
Then algorithms and software are developed and, in some cases, the analysis results in superior performance as judged by figures-of-merit applied to that data set, which leads to a publication.
Are the results extendible and do they have `error bars'?

Usually figures-of-merit are concerned with the \textit{efficiency} of a method.
However, \textit{validity} should be established first, and error bars allow this to be done.
No matter how efficient a prediction method is, if its error bars define a nominal 90\% prediction interval but the true quantity (i.e., the predictand) is only contained in these intervals 60\% of the time, the method lacks validity.
This important figure-of-merit is called \textit{coverage} and is analogous to one in the medical literature called specificity.

This article is aimed at data scientists who want their error bars to be based on probabilities and their coverages to be (approximately) as stated.
There are data scientists who have error bars that are based on empirical distributions derived from ensembles of like data sets or from partitioning a single data set into like subsets.
While the principles I present are for data scientists who quantify uncertainties using probabilities, they could act as a guide for both types.
If your analyses include uncertainty quantifications, then you may find some of the principles in the following sections useful for hypothesis generation, for climbing the knowledge pyramid from data to information to knowledge, and for scientific inference.

\section{My principal principles}

When I was planning this article, I hovered between the words `laws,' `rules,' `principles,' and `guidelines' to establish some landmarks in Data Science.
Laws should not be broken, exceptions prove the rule, and guidelines prevent accidents.
But principles should make the analysis go better without interdicting other approaches: Improvising on something Groucho Marx once said, if you don't like my principles, you may have others you like better $\ldots$

A data scientist might see variability between strata and within strata.
A statistical scientist uses probability distributions to capture some or all of this variability. 
For example, probability densities would typically be used to model within-strata variabilities, and these densities may be different across strata in order to capture the between-strata variability.
In many, many cases, those strata differences are modelled via a regression model, leaving the errors around the regression line (i.e., the within-strata variability) to be described by a probability distribution, often a Gaussian (or sometimes called normal) distribution.

In my opinion, the worst word in the Statistics lexicon, which happens to have four letters, is `mean.'
It is the first moment of a probability distribution, and it is also used to describe the average of a collection of numbers, leading to many misunderstandings in Data Science.
To data scientists who capture variability through the empirical distributions of summary statistics from similar data sets, the average is their first moment, so \textit{they} could legitimately call it a mean of their empirical distribution.
To avoid confusion, it will be henceforth assumed in this article that variability is captured by \textit{probability} distributions, and a mean \textit{iss} (is and only is) the first moment of that distribution.

In this section, five general statistical principles are presented.
Different areas of study will have their own and, in Section 3, I present three principles that are important in Spatial Statistics.
In an epilogical section (Section 4), I present one more, the principle of respecting units, which makes nine principles in total.

\subsection{All models are wrong $\ldots$ but some are wrong-er than others}

In my work on remote sensing of carbon dioxide, data $\bz$ from a single sounding is a more-than-2,000-dimensional vector of radiances observed at wavelengths in the near infra-red part of the spectrum.
\citet{cressie2018} contains a literature review that features the relationship between $\bz$ and the state of the atmosphere $\by$ (a 40-to-50 dimensional vector) via the statistical model,
\begin{equation}
\label{eq:model}
\bz=\boldf(\by)+\bfvarepsilon\,,
\end{equation}
where unobserved $\by$ and $\bfvarepsilon$ are statistically independent; $\bfvarepsilon\sim\textrm{N}(\bzero,\bSigma_\varepsilon)$, where $\textrm{N}(\bfmu,\bSigma)$ refers to a multivariate Gaussian distribution with mean vector $\bfmu$ and covariance matrix $\bSigma$; $\boldf$ is a known vector of nonlinear forward models obtained from atmospheric physics/chemistry; $\bSigma_\varepsilon$ is a covariance matrix known from experiments on the remote sensing instrument, performed in the laboratory before the satellite is launched; $\by=\bfmu_\alpha+\bfalpha$, where $\bfalpha\sim\textrm{N}(\bzero,\bSigma_\alpha)$; and $\bfmu_\alpha$ and $\bSigma_\alpha$ are the first two moments of the (hidden) atmospheric state $\by$.
The goal is to predict $\by$ from data $\bz$.
In my experience, it is the specification of $\bSigma_\alpha$ (the covariance matrix of the state of the atmosphere) where the statistical model can be seriously wrong~\citep{cressie2018}.

\noindent\textbf{Principle 2.1}: \textit{Establish a true model (TM), perhaps different from the scientist's working model (WM). Critically, compute the TM-distributional properties of the WM estimators.}

At the very least, this principle gives the data scientist an idea of the sensitivity of the model to misspecification.
In the remote sensing example, the WM-based statistical analysis provides a predicted atmospheric state, $\hat{\by}_{W\!M}$, called a \textit{retrieval.}
However, the atmosphere does not care what WM was chosen; it acts under the true model (TM), and the principal figures-of-merit that are calculated are (using obvious notation):
\begin{equation}
\begin{aligned}
\text{True bias: }&\mathrm{E}_{T\!M}(\hat{\by}_{W\!M}-\by)\,,\label{eq-ETM}\\
\text{True uncertainty: }&\var_{T\!M}(\hat{\by}_{W\!M}-\by)\,.
\end{aligned}
\end{equation}
While $\hat{\by}_{W\!M}$ may be optimal under WM, it is not under TM, and the properties given by \eqref{eq-ETM} will expose the potential weaknesses of a WM chosen for its simplicity or its computational convenience.

It is tempting (but wrong) for scientists to calculate instead:
\begin{equation}
\begin{aligned}
&\mathrm{E}_{W\!M}(\hat{\by}_{W\!M}-\by)\,,\label{eq-EWM}\\
&\var_{W\!M}(\hat{\by}_{W\!M}-\by)\,,
\end{aligned}
\end{equation}
which are often facile calculations compared to \eqref{eq-ETM}.
This was the case for NASA's Orbiting Carbon Observatory-2 (OCO-2) satellite.
Over time, empirical distributions of retrieval-error ensembles showed that bias and variance should have been calculated from \eqref{eq-ETM}, not \eqref{eq-EWM}.
This is discussed by \cite{cressie2018}, and an illustration of applying Principal 2.1 is given in \cite{nguyen2019}.

In the rest of this article, I shall assume that the WM and TM are one and the same.
However, this principle is potentially applicable in all the examples below.

\subsection{What you see is not what you want to get}

There is an equivalent way to write the model \eqref{eq:model} that shows how statistical scientists can naturally build complexity into their models:
\begin{equation}
\begin{aligned}
\bz|\by&\sim\textrm{N}(\boldf(\by),\bSigma_\varepsilon)\label{eq:z|y}\\
\by&\sim\textrm{N}(\bfmu_\alpha,\bSigma_\alpha)\,.
\end{aligned}
\end{equation}
This is a \textit{hierarchical statistical model (HM)}, which is defined by layers of conditional probabilities (here two layers).

The top layer is the \textit{data model} and can be written in abbreviated notation as $[\bz|\by]$.
It models what we see -- the data (conditional on the process $\by$).
The next layer is the \textit{process model} that can be written as $[\by]$.
It models what we want to get -- the latent process, or state, hidden behind the data.
In fact, in the remote sensing problem described in Section 2.1, the distribution of $\by$ is conditional on $\bw$, the meteorology of the atmosphere.
If need be, the process model could be modified to now consist of $[\by|\bw]$ and $[\bw]$.
Then the HM would be made up of three levels, and the sequence of conditional probabilities would be $[\bz|\by]$, $[\by|\bw]$, and $[\bw]$.
In the specific case of OCO-2 retrievals, the meteorological process, $\bw$, is obtained from a numerical weather forecasting model and is considered known.

\noindent\textbf{Principle 2.2}: \textit{Build statistical models conditionally, through a data model and a process model. Infer the unknown process from the predictive distribution.}

The predictive distribution is,
\begin{equation}
\label{eq:pred-dist}
[\by|\bz]=\frac{[\bz|\by]\cdot [\by]}{[\bz]}\,,
\end{equation}
where for convenience any parameters $\thetab$ in $[\bz|\by]$ and $[\by]$ are dropped from the notation (but they are still there).
The HM components are featured in the numerator, which is equal to the joint distribution, $[\bz,\by]$; and the denominator, $[\bz]$, is simply a `normaliser' to ensure that $[\by|\bz]$ is a density (or probability mass) function that integrates (or sums) to $1$.
Equation \eqref{eq:pred-dist} is Bayes' Theorem but, in this case, the unknowns are the state elements in $\by$, not the parameters $\thetab$.
No prior distributions on $\thetab$ have been assumed in the making of this HM!
However, a prior could be assumed or, pragmatically, an estimate $\hat{\thetab}$ could be obtained from $\bz$.

Derivation of the predictive distribution in \eqref{eq:pred-dist} can be problematic in many cases of practical interest, such as when $[\by]$ is a highly multivariate probability distribution of a complex scientific process.
Computational methods, particularly Markov chain Monte Carlo (MCMC), are in constant development to allow realisations from $[\by|\bz]$ to be simulated, from which distributional properties such as the predictive mean and the predictive quantiles can be used to predict the state $\by$ and the error bars, respectively.

The simplest special case of~\eqref{eq:z|y} is when the data (here, $z$) and the process (here, $y$) are univariate, and $f(\cdot)$ is the identity function:
\begin{align*}
z|y&\sim\textrm{N}(y,\sigma_\varepsilon^2)\,,\\
y&\sim\textrm{N}(\mu_\alpha,\sigma_\alpha^2)\,.
\end{align*}
Then $[y|z]$ is Gaussian, characterised by its first two moments:
\begin{align*}
\mathrm{E}(y|z)&=\{(1/\sigma_\alpha^2)\mu_\alpha+(1/\sigma_\varepsilon^2)z\}\{1/\sigma_\alpha^2+1/\sigma_\varepsilon^2\}^{-1}\,,\\
\var(y|z)&=\{1/\sigma_\alpha^2+1/\sigma_\varepsilon^2\}^{-1}\,.
\end{align*}
The predictor, $\hat{y}=\mathrm{E}(y|z)$, is a weighted combination of the data and the mean of the unknown state $y$, where the weights depend on the signal-to-noise ratio, $\sigma_\alpha^2/\sigma_\varepsilon^2$.
These results can be generalised to the multivariate case with linear retrieval equations in \eqref{eq:z|y}, namely $\boldf(\by)=\bc+\bK\by$.
In the case of OCO-2, $\bK$ has more rows than columns, and the first two moments of the predictive distribution, which is Gaussian, are:
\begin{equation}
\begin{aligned}
\label{eq:data-z}
\mathrm{E}(\by|\bz)&=\by_\alpha+\bG(\bz-\bc-\bK\by_\alpha)\,,\\
\var(\by|\bz)&=\{\bSigma_\alpha^{-1}+\bK^\top\bSigma_\varepsilon^{-1}\bK\}^{-1}\,,
\end{aligned}
\end{equation}
where $\bG=\{\bSigma_\alpha^{-1}+\bK^\top\bSigma_\varepsilon^{-1}\bK\}^{-1}\bK^\top\bSigma_\varepsilon^{-1}$ is sometimes called the gain matrix.

From \eqref{eq:data-z}, it can be seen that the precision matrix (i.e., the inverse of the variance matrix), written $\text{prec}(\cdot)$, of each component of the predictive distribution satisfies,
$$
\text{prec}(\by|\bz)=\text{prec}(\by)+\bK^\top\text{prec}(\bz|\by)\bK\,,
$$
a result that holds when the matrix $\bK$ is any size.
This decomposition of precision demonstrates that, when going from $[\by]$ to $[\by|\bz]$, the precision increases (i.e., the variance decreases). 
Moreover, the predictive mean is unbiased; that is, $\mathrm{E}(\mathrm{E}(\by|\bz))=\mathrm{E}(\by)=\bfmu_\alpha$.

In geostatistics, the importance of \eqref{eq:pred-dist} is deeply misunderstood, because \cite{matheron1963} originally formulated kriging in terms of what he called a regionalised variable, $\{z(\bs):\bs\in D\subset \RR^d\}$, where $\RR^d$ is $d$-dimensional Euclidean space and $D$ is a subset of $\RR^d$ with volume $|D|>0$.
In this formulation, the data model and the process model are collapsed into the single probability distribution, $[\{z(\bs):\bs\in D\}]$.
Then the goal is to predict $z(\bs_0)$ from data $\bz=(z(\bs_1),\ldots,\bz(\bs_n))^\top$, for which the generic kriging predictor is $\mathrm{E}(z(\bs_0)|\bz)$, the mean of $[z(\bs_0)|\bz]$.

However, when measurement of the process is taken into account, there is a HM that differentiates between the observations $\bz$ and the underlying latent process $\{y(\bs):\bs\in D\}$ and that is observed imperfectly through $\bz$. % a data model, $[\bz|\{y(\bs):\bs\in D\}]$, and a process model, $[\{y(\bs):\bs\in D\}]$.
The spatial trend and the spatial covariance function are defined in the process model, although they are estimated from the noisy data $\bz$.
Once the measurement error is included in the model, it is clear that geostatistics should do kriging using $\mathrm{E}(y(\bs_0)|\bz)$ and not $\mathrm{E}(z(\bs_0)|\bz)$; see \citet[Section 4.1]{cressie2011}.

Consequently, much of the earlier geostatistics software did not make inference on $y(\bs_0)$ but chose to make inference on $z(\bs_0)$ where the measurement error is confounded with the process error.
User beware: some still do, but those written for environmental applications (e.g., \textsf{geoR}, \textsf{gstat}, \textsf{FRK}) give the correct kriging equations.
The difference is most apparent when the kriging variance is computed as $\var(z(\bs_0)|\bz)$ at location $\bs_0$, but then interpreted incorrectly as the predictive variance $\var(y(\bs_0)|\bz)$ of the true process at $\bs_0$.

\subsection{Geophysicists conserve energy but what do data scientists conserve?}

Building physical models usually involves ensuring that mass or energy is conserved.
If the system is leaking energy, then it needs to be plugged or the energy needs to be followed as it moves into another system.

Now consider a designed experiment where data $\bz$ are obtained and the statistical model fitted is
\begin{equation}
\label{eq:bz}
\bz=\bX\bbeta+\bxi\,.
\end{equation}
The model in \eqref{eq:bz} is a linear regression with covariate (or design) matrix $\bX$, $\bbeta$ is an unknown $p$-dimensional vector of regression coefficients, and $\bxi=(\xi_1,\ldots,\xi_n)^\top$ consists of random variables that are independent and identically distributed (iid) Gaussian random variables with mean $0$ and variance $\sigma_\xi^2$.

Suppose that the measuring instrument was carefully calibrated and, in the study protocol, its sample variance from repeated measurements was reported as a number that I shall write as $s_\varepsilon^2>0$.
Note that the uncertainty in the measurements usually needs to be quantified outside the experiment in order to identify the linear-model-error variance.
Scientific interest is primarily in $\bbeta$, but $\sigma_\xi^2$ is by no means a nuisance parameter.
Its restricted maximum likelihood (REML) estimate is:
$$
s_\xi^2=(\bz-\bX\hat{\bbeta})^\top(\bz-\bX\hat{\bbeta})/(n-p)\,,
$$
where $\hat{\bbeta}$ is the maximum likelihood (and also the ordinary least squares) estimate of $\bbeta$.

It could almost be a rule that in any study of this sort, you will see $s_\xi^2>s_\varepsilon^2$. 
Where did the component of variability, ($s_\xi^2-s_\varepsilon^2$), go?

\noindent\textbf{Principle 2.3}: \textit{In any well defined statistical model, there is conservation of variability.}

\noindent Indeed, the model given by \eqref{eq:bz} is not defined well enough: The error term $\xi_i$ is, in fact, made up of two components of variance:
$$
\xi_i=\delta_i+\varepsilon_i\,;\,i=1,\ldots,n\,,
$$
where $\{\varepsilon_i:i=1,\ldots,n\}$ represent measurement errors.
Often forgotten are $\{\delta_i:i=1,\ldots,n\}$, which represent model errors resulting from using the linear-regression model, $\{\bx_i^\top\bbeta:i=1,\ldots,n\}$, and which are key in accounting for the inexactness of using any model (linear or nonlinear).

The scientific process $\{y_i:i=1,\ldots,n\}$ is given by $y_i=\bx_i\bbeta+\delta_i$, and an observation of it is $z_i=y_i+\varepsilon_i$, for $i=1,\ldots,n$.
The HM captures the variability of $\bz$ and $\by$ beautifully, as follows:
\begin{equation}
\begin{aligned}
\label{eq:z|y-model}
\bz|\by&\sim\textrm{N}(\by,\sigma_\varepsilon^2\bI)\,\,[\text{or, }\bz=\by+\bfvarepsilon]\,,\\
\by&\sim\textrm{N}(\bX\bbeta,\sigma_\delta^2\bI)\,\,[\text{or, }\by=\bX\bbeta+\bfdelta]\,.
\end{aligned}
\end{equation}
Hence, from \eqref{eq:z|y-model}, we obtain the earlier result in vector form:
\begin{equation}
\label{eq:bz-model}
\bz=(\bX\bbeta+\bfdelta)+\bfvarepsilon=\bX\bbeta+(\bfdelta+\bfvarepsilon)=\bX\bbeta+\bxi\,,
\end{equation}
where $\bfdelta$ and $\bfvarepsilon$ are independent mean-zero random vectors.

Comparing \eqref{eq:bz} and \eqref{eq:z|y-model}, we see that the model given by \eqref{eq:bz} should be augmented with the equation, $\bxi=\bfdelta+\bfvarepsilon$, which results in the conservation-of-variability equation, $\sigma_\xi^2=\sigma_\delta^2+\sigma_\varepsilon^2$.
Consequently, knowing $s_\varepsilon^2$ (an estimate of $\sigma_\varepsilon^2$ that is obtained outside the experiment) and, having computed the REML estimate of $\sigma_\varepsilon^2$, an estimate of the model error, $\sigma_\delta^2$, can be obtained as,
\begin{equation}
\label{eq:s-delta}
s_\delta^2= s_\xi^2-s_\varepsilon^2
\end{equation}
(provided the right-hand side is non-negative).

In its simplest form, conservation of variability says that the \textit{total variability} is equal to the variability due to \textit{model uncertainty plus} the variability due to \textit{measurement uncertainty}.
When variability is captured using a probability space defined by a HM, this principle can be expressed as:
\begin{equation}
\label{eq:var-xi}
\var(\bz)=\var(\mathrm{E}(\bz|\by))+\mathrm{E}(\var(\bz|\by))\,.
\end{equation}
For example, consider the HM defined by \eqref{eq:z|y-model}.
Since $\var(\bz)=\var(\bxi)$ and $\var(\by)=\var(\bfdelta)$, we have
$$
\sigma_\xi^2\bI=\sigma_\delta^2\bI+\sigma_\varepsilon^2\bI\,,
$$
and variability is conserved.

This might seem obvious to you, or perhaps even trivial.
Again consider \eqref{eq:bz}, and suppose you want to predict an unknown value, $y_{n+1}$, outside the data set, but only the $p$-dimensional estimate $\hat{\bbeta}$ and the covariate, $\bx_{n+1}$, are at your disposal.
Most regression textbooks would say that you should use as predictor, $\bx_{n+1}^\top\hat{\bbeta}$.
However, a scientist wants to predict the value of the scientific process, $y_{n+1}=\bx_{n+1}^\top\bbeta+\delta_{n+1}$.
Using the well defined HM \eqref{eq:z|y-model}, its predictor is $\hat{y}_{n+1}=\mathrm{E}(y_{n+1}|\bz)=\bx^\top_{n+1}\hat{\bbeta}+\hat{\delta}_{n+1}$, where $\hat{\delta}_{n+1}= \mathrm{E}(\delta_{n+1}|\bz)$.
Since $\{\delta_i\}$ are iid $\textrm{N}(0,\sigma_\delta^2)$, $\hat{\delta}_{n+1}= \mathrm{E}(\delta_{n+1}|\bz)=0$; however, $\var(\delta_{n+1}|\bz)$ is \textit{not zero}, and that has to be recognised in order to perform valid predictive inference, as given below~\citep[e.g.,][]{cressie2020}.

The prediction error is $(\hat{y}_{n+1}-y_{n+1})$, and its first two moments are:
\begin{align*}
&\mathrm{E}(\hat{y}_{n+1}-y_{n+1})=0\,,\\
&\mathrm{E}(\hat{y}_{n+1}-y_{n+1})^2=\mathrm{E}(\bx_{n+1}^\top(\hat{\bbeta}-\bbeta))^2+\mathrm{E}(\var(\delta_{n+1}|\bz))\,,
\end{align*}
since the expectation of the cross-product is zero.
Thus, if $\var(\delta_{n+1}|\bz)$ were forgotten by the data scientist, such as would happen if the presence of $\{\delta_i\}$ were not recognised in the statistical model, a forecasting decision about future values of the process $\{y_i\}$ would be overly optimistic and potentially harmful.

Principle 2.3 also shows up in the analysis of variance (ANOVA) method, where the `between sum of squares' corresponds to the variance of the conditional expectation in \eqref{eq:var-xi}, the `within sum of squares' corresponds to the expectation of the conditional variance in \eqref{eq:var-xi}, and the `total sum of squares' corresponds to the left-hand side of \eqref{eq:var-xi}.
Conservation of variability implicitly includes variability and \textit{covariability}.
For example, if two random variables $\varepsilon_1$ and $\varepsilon_2$ have correlation $\rho\neq0$, then it should be recognised that $\var(\varepsilon_1+\varepsilon_2)\neq\var(\varepsilon_1)+\var(\varepsilon_2)$
This has often been ignored by scientists doing an error budget (e.g., in remote sensing retrievals; see \citealp{connor}).
Depending on the sign of $\rho$, $\var(\varepsilon_1)+\var(\varepsilon_2)$, may be either larger than or smaller than the total variability, since $\var(\varepsilon_1+\varepsilon_2)=\var(\varepsilon_1)+\var(\varepsilon_2)+2\rho\var(\varepsilon_1)^{1/2}\var(\varepsilon_2)^{1/2}$.

The rules of probability theory can explain easily how the variability can seem to disappear, and then they can show us where to find it.
It is less natural to do so simply with empirical distributions, because the pairs $\{(z_i,y_i)\}$ involve the unavailable, hidden variable $\{y_i\}$.
Further, if $\{z_i\}$ and $\{x_j\}$ are two sets of observations, there is no guarantee that they will occur in pairs, and hence the empirical correlation, $r$, might be difficult to obtain.

The famous bias--variance trade-off is another manifestation of Principle 2.3.
In the context of estimation of a fixed but unknown parameter $\theta$ with an estimator $\hat{\theta}(\bz)$, where recall $\bz$ is the data vector, the mean-squared error is the sum of the estimator's squared bias and its variance:
$$
\mathrm{E}(\hat{\theta}(\bz)-\theta)^2=(\mathrm{E}(\hat{\theta}(\bz))-\theta)^2+\var(\hat{\theta}(\bz))\,.
$$
Generally speaking, an estimator that decreases the bias will increase the variance, and \textit{vice versa}.
The mean-squared error might be decreased using different estimators, but there is a trade-off to be made when trying to decrease both bias and variance at the same time.

\subsection{The holy grail: all scales of variation are additive}

A statistical analysis cannot get very much simpler than fitting a simple linear regression (a special case of \eqref{eq:bz}) to $\{(z_i,x_i):i=1,\ldots,n\}$, as follows:
\begin{equation}
\label{eq:z-i}
z_i=\beta_1+\beta_2 x_i +\xi_i\,;\,i=1,\ldots,n\,,
\end{equation}
where $\{\xi_i:i=1,\ldots,n\}$ are iid $\textrm{N}(0,\sigma_\xi^2)$.
However, is \eqref{eq:z-i} appropriate if $\{z_i:i=1,\ldots,n\}$ are photon counts at wavelengths in a given band of the electro-magnetic spectrum, or if they are percentages of trace elements in soil?

Exploratory data analysis (EDA) might reveal a highly skewed histogram of $\{z_i\}$.
After plotting the histogram of $\{\log z_i\}$, we might achieve a more symmetric appearance; assume for the moment that we do.
This would prompt a plot of not only $\{z_i\}$ versus $\{x_i\}$, but also of $\{\log z_i\}$ versus $\{x_i\}$, and the latter might look more linear.
Further EDA, in the form of residual plots, after fitting a simple linear regression to the data $\{z_i\}$ and one to the transformed $\{\log z_i\}$, would be carried out.
And, for example, it might be that the residual variability of $\{z_i\}$ appears to increase with $x$ (i.e., heteroskedasticity), but the residual variability of $\{\log z_i\}$ shows no dependence on $x$ (i.e., homoskedasticity).

Admittedly, this is a story, not methodology, but in my experience with analysing data, it happens enough to propose the next principle~\citep{cressie1985}.

\noindent\textbf{Principle 2.4}: \textit{Seek a transformation of the scientific process where all components of variation act and interact additively.}

\noindent This principle looks more like a doctrine but, as I explain below, a fitting of \eqref{eq:z-i} to the data $\{z_i\}$ is following this principle, where the transformation is simply the identity.

Suppose the quantity $y$ has variabilities that can be expressed as `large-scale' and `small-scale.'
Scientists usually put their theories in the large scale and their errors (both in the theory and in the measurements) in the small scale.
Statistical scientists know that to make inference on the large scale of the scientific process, its small scale has to be modelled well, usually by a random error term, $\delta$.
However, the model I now give for the scientific process is more basic, given in terms of those large and small scales of variability:
It is additive and of the following form,
\begin{equation}
\label{eq:y}
y=(\mu^{(1)}+\ldots+\mu^{(p)})+(\delta^{(1)}+\ldots+\delta^{(N)})\,,
\end{equation}
where $p$ and $N$ are positive integers and, for convenience, the subscript `$i$' has been dropped.
For example, the large-scale variation in simple linear regression has $p=2$, $\mu^{(1)}=\beta_1$, and $\mu^{(2)}=\beta_2 x$.
For the small-scale variation in \eqref{eq:y}, $N$ is large, and $\{\delta^{(l)}:l=1,\ldots,N\}$ are physically interpretable components with variabilities that are very small but such that their total variability is the variability of the difference, $y-\mu^{(1)}-\ldots-\mu^{(p)}$; see Principle 2.3.

The model \eqref{eq:y} is a physical model where the large scales act additively with each other, the small scales act additively with each other, and the large scales and the small scales interact additively.
It can be made statistical by assuming that these many small-scale effects, $\delta^{(1)},\ldots,\delta^{(N)}$, are random, statistically independent, have mean zero and variances that are $O(1/N)$, with the same leading coefficient, denoted here as $\sigma_\delta^2$.
Now let $N$ be large and define the small-scale variations collectively as $\delta$; then \eqref{eq:y} can be written as, 
$$
y=\mu^{(1)}+\ldots+\mu^{(p)}+\delta\,,
$$
where
\begin{enumerate}
\item $\mathrm{E}(y)=\mu^{(1)}+\ldots+\mu^{(p)}$ (additivity)\,,
\item $\var(y)=\sigma_\delta^2$ (homoskedasticity)\,,
\item $\delta \sim \textrm{N}(0,\sigma_\delta^2)$ (central limit theorem)\,.
\end{enumerate}

The regression \eqref{eq:bz} is obtained when the large-scale effects $\{\mu^{(k)}:k=1,\ldots,p\}$ are identified with $\{\beta_k x_k:k=1,\ldots,p\}$.
It is the default model used in much of science, but this discussion shows that it originates from the imposition of additivity within and between scales of variability.
From Principle 2.2, the measurement of a phenomenon should be separated from how the phenomenon behaves in nature and, indeed, the generalised linear model does this through a link function~\citep{mccullagh1989}.
Hence, Principle 2.4 covers many statistical models and offers a plausible explanation of why a linear relation, homoskedasticity, and Gaussianity are often found to occur together after a transformation of the data~\citep[e.g.,][]{cressie1978}.
Of course, it is easy to construct probability models where the principle does not hold, but one should keep in mind that probability theory is used to model nature's variability, not the other way around.

When nonlinearity is inherent to the physical system, such as would occur when there are barriers or thresholds, the quest looks to be futile.
However, after following Principle 2.2, it may just be that the grail, Principle 2.4, is hidden deep in the process model~\citep{berliner2000}.

\subsection{Could swans be red?}

\cite{taleb2007} published a book about `Black Swan' events that, when they happen, are considered to have been unpredictable.
Such events could have a major impact on Earth's environment or, as has happened in 2020, on global-population health and economies around the globe.

This black-swan meme has its roots in seventeenth-century Europe.
Up to then, no swans had ever been observed that were black but, in 1697, Dutch explorers became the first Europeans to observe black swans during a voyage that took them to Western Australia.
The implication in Taleb's book is that if the scientific world in the 1600s could not predict black swans, how can scientists predict catastrophic environmental events in the twenty-first century?

My response is that 300+ years ago the best scientific minds in Europe were too certain about their science; that is, if they had been asked to put probabilities (according to abundance) on the colours that swans could be: red, orange, $\ldots$, violet, white, black, they would have put $0,0,\ldots,1,0$, which is a well defined probability mass function.
Certainly, black-swan events are \textit{not} predictable if the scientific model is 100\% certain that they do not exist.
Because their probability model gave black swans and indeed coloured (including red) swans, zero probability, this unimagined event did not emerge out of the `ether' in subsequent inferences, until one was observed.
Do we have to wait until a highly unusual event occurs before we are forced to change the probability model?
The real lesson from the black-swan meme is that scientific knowledge is never perfect, that modellers need to explore the parameter space thoroughly, and that they need to `spend' some probability on highly unusual events.

At the time of writing this article, our species is under attack from a virus that was originally called `novel coronavirus,' so new that it took several weeks before the infection it caused had a name: COVID-19.
Virologists certainly had assigned a small positive probability that each new decade would have its own `novel' pandemic.
However, politicians (and most economists) appear to have put zero probability on a severe, worldwide economic disruption.

Given a swine-flu-like pandemic has occurred, the conditional probability of some economic disruption is \textit{not zero}.
But, given a severe pandemic has occurred, the conditional probability of severe economic disruption is substantial.
This conditional probability is then multiplied by the probability of a severe pandemic, which is by no means zero given the ability of viruses to mutate and occasionally jump from animals to humans. 
The product of these two is the joint probability of a severe pandemic followed by severe economic disruption, which is not negligible.
This has happened twice in the last hundred years, and it will likely happen more often with humans and animals sharing in an ever-more-crowded environment.

\noindent\textbf{Principle 2.5}: \textit{When building probability models, look carefully where zero probabilities are assumed (perhaps implicitly) and, with the same care, move appropriate probabilities away from zero.
Calculate joint probabilities from products of conditional (not marginal) probabilities, unless entropy is maximal.}

Unfortunately, uncertainty quantification through joint probabilities all too often comes from multiplying marginal probabilities as if each event in the causal chain were independent.
It does not `hurt' for small probabilities to be assigned to Pr(bird is colour $c$ $\vert$ bird is a swan) where $c$ also covers the colour `black.'
Then Pr(bird is a swan and it is black) = Pr(bird is black $\vert$ bird is a swan) $\times$ Pr(bird is a swan).
The point being made here is philosophical, and it could be applied to Pr(pandemic followed by global economic disruption), for different severities of both events.
The worst thing would be to make Pr(global economic disruption $\vert$ pandemic) equal to zero, since then there would be a lack of planning for the health and economic crises brought on by a pandemic~\citep{mackenzie2020}.

If nothing were known about the conditional probability, a fall-back is the maximum-entropy model where the marginal probability, Pr(bird is colour $c$), is used in place of Pr(bird is colour $c$ $\vert$ bird is a swan). %~\citep{cressie2004}.
The marginal probability is not zero, since it is based on abundance and, for example, there are birds that are predominantly coloured red (e.g., Australia's king parrot).
The maximum-entropy principle is discussed in~\citet{cressie2004}. 

Experiences over the last 100 years mean that Pr(global economic disruption), whether caused by a pandemic or not, should not be zero, yet governments described the events of March--April 2020 as unimaginable.
Events of small (but not zero) probabilities with consequences expressed through a loss function, allow a non-zero expected loss to be calculated, which can then be used to make optimal decisions that mitigate the consequences (e.g., see~\citealp{berger1985} for a discussion of decision theory).

I have just presented five statistical principles that should be useful in Data Science.
Data often come with location information, in which case the data scientist will likely be using spatial-analysis methods.
In the next section, I present three principles that are specific to Spatial Statistics.

\section{A few spatial statistical principles}

Those of us who work in Spatial Statistics will know of Tobler's First Law of Geography \citep{tobler1970}.
In spatial statistical science, it really is a `principle' rather than a `law' and, in the following subsections, I shall present it and two other principles that I have found useful in this area of research.

\subsection{Patches in close proximity are commonly more alike $\ldots$}

In his famous 1935 book on experimental design, \citet[][p. 66]{fisher1935}, wrote: `After choosing the area we usually have no guidance beyond the widely verified fact that patches in close proximity are commonly more alike, as judged by the yield of crops, than those which are further apart.'
A spatial statistician sees this as the making of a principle but, at that time, Fisher made a sharp right turn.
In his analyses of field trials he applied randomisation to eradicate the pest: spatial correlation!
\citet[Ch.1 and Ch. 4]{cressie2011} give some historical perspective to the work of researchers who took roads less travelled and developed this vibrant area we now call Spatial Statistics.
Some of this development comes from Geography, and so it is fitting that the first and most important principle in this section has become known as the First Law of Geography.
Originally articulated by \cite{tobler1970}, it is given here in exactly his words.

\noindent\textbf{Principle 3.1} \textit{Everything is related to everything else, but near things are more related than distant things~\citep{tobler1970}.}

This principle is at the core of what we do in spatial and spatio-temporal statistics.
For example, in remote sensing of Earth's surface, the scene of interest $D$ contains \textit{many} retrievals, $\{z(\bs_i):i=1,\ldots,n\}$, where $\{\bs_i:i=1,\ldots,n\}$ are the (lon, lat) locations of the $n$ data inside $D$ (retrieved over a short time period).
There is a hidden spatial process $\{y(\bs):\bs\in D\}$ that the geophysicist would like to infer and, in a spatial HM, a spatial statistical model for $\{y(\bs):\bs\in D\}$ is built around Principle 3.1.

For example, consider a simple process model, appropriate for a small scene $D$: 
$$
y(\bs)=\beta_1+\beta_2\text{lat}(\bs)+\delta(\bs)\,,\text{ for }\bs\in D\,,
$$
where the mean of the process is a linear function of latitude, and $\{\delta(\bs):\bs\in D\}$ is a spatial error process with mean zero and stationary covariance function, $C_\delta(\bh)=\cov(\delta(\bs+\bh),\delta(\bs))=\sigma_\delta^2\cdot\text{exp}(-\|\bh\|/\phi)$.
Notice that $C_y(\bh)=\cov(y(\bs+\bh),y(\bs))=C_\delta(\bh)$; $C_y(\bzero)=C_\delta(\bzero)=\sigma_\delta^2$; and the scale parameter $\phi$ controls how `more related' things are.
With this parameterisation, $\phi=0$ is the degenerate case of no spatial correlation and, as $\phi$ increases, the spatial correlation increases for a given distance $\|\bh\|$.
The data model in this example is also simple, that of additive independent measurement error.
The data vector is $\bz=(z(\bs_1),\ldots,z(\bs_n))^\top$ and
$$
z(\bs_i)=y(\bs_i)+\varepsilon_i\,;\,i=1,\ldots,n\,,
$$
where $\{\varepsilon_i:i=1,\ldots,n\}$ are independent mean-zero errors.

Assuming for the moment that all the parameters $\thetab=(\bbeta^\top,\sigma_\delta^2,\phi,\sigma_\varepsilon^2)^\top$ are known, inference on $\{y(\bs):\bs\in D\}$ comes from summaries of the predictive distribution, $[\{y(\bs):\bs\in D\}|\bz]$.
For a Gaussian process model $y(\cdot)$ and Gaussian measurement errors $\{\varepsilon_i\}$, the predictive distribution is also a Gaussian process whose first two moments, $\mathrm{E}(y(\bs)|\bz)\text{ and }\cov(y(\bs),y(\bu)|\bz)$, for $\bs,\bu\in D$, can be obtained analytically.
From Section 2.2, a common predictor of $y(\bs)$ is $\hat{y}(\bs)= \mathrm{E}(y(\bs)|\bz)$, whose statistical properties are needed for inference.
It is straightforward to see that $\mathrm{E}(\hat{y}(\bs))=\mathrm{E}(y(\bs))$ (i.e., the predictor is unbiased) and the mean-squared predictor error is:
$$
\mathrm{E}(\hat{y}(\bs)-y(\bs))^2=\mathrm{E}(\var(y(\bs)|\bz))=\var(y(\bs)|\bz)\,,
$$
which is the predictive variance (the last equality being due to the Gaussian assumptions made).
Further summaries might come from well chosen percentiles (e.g., 2.5\%, 25\%, 50\%, 75\%, 97.5\%) of $[y(\bs)|\bz]$.

\subsection{What is one person's mean function could be another person's spatial error}

In spatial statistics, it might appear that the spatial-prediction problem is quite difficult, because the number of predictions to be made is often greater than the number of data.
Even for the problem of parameter estimation, the number of `degrees of freedom' in $n$ spatially dependent data will not be $n$, as I now show.
Applying Principle 3.1 with a stationary spatial covariance model $C_z(\bh)=\cov(z(\bs+\bh),z(\bs))$, for $\bh\in\RR^d$, that exhibits only positive correlations, it is easy to see that for $\overline{z}=(1/n)\sum_{i=1}^n z(\bs_i)$,
$$
\var(\overline{z})=(1/n)\{\sigma_z^2+2\mathop{\sum\sum}_{i<j} C_z(\bs_i - \bs_j)/n\}>\frac{\sigma_z^2}{n}\,,
$$
where $\sigma_z^2= C_z(\bzero)$.
Hence one can define the \textit{effective degrees of freedom}, $n_{\text{eff}}$, as
$$
n_{\text{eff}}=\sigma_z^2/\var(\overline{z})<n\,;
$$
that is, under the almost ubiquitous spatial model of positive spatial correlation, $n_{\text{eff}}<n$.
In the remote sensing context where there are many observations taken over a short period of time,~\citet{zhang2017} gave an example where $n=2961$ and $n_{\text{eff}}=202.3$, less than one-tenth of the sample size!

These calculations are a warning that, in the spatial (and spatio-temporal) setting, intuition learned from the `iid errors' model has to be modified, sometimes substantially.
Critically, the probability model that captures the spatial variability has to be well specified, and we now discuss how this can be done.

The classical spatial statistical model consists of a deterministic mean process, $\{\mu(\bs):\bs\in D\}$, and a random spatial error process, $\delta(\bs)= y(\bs)-\mu(\bs)$, so that
\begin{equation}
\label{eqn-y-bs}
y(\bs)=\mu(\bs)+\delta(\bs)\,,\text{ for }\bs\in D\,;
\end{equation}
see~\citet[Ch. 2]{cressie1993}.
It is often a personal choice what components go into $\mu(\bs)$.
For linear regression, $\mu(\bs)=\bx(\bs)^\top\bbeta$, for $\bs\in D$, but the set of possible covariates, $\bx(\bs)$, typically comes from the modeller's subject-matter knowledge, augmented by \textit{spatial trend} terms that are linear and higher-order functions of $\bs=(s_1,\ldots,s_d)^\top$.

What is the effect of not having included an important spatially varying covariate, $x_{p+1}(\bs)$ say, in $\mu(\bs)$?
From Principle 2.3, $\{\delta(\bs)\}$ has to absorb the contribution to spatial variability that $\{x_{p+1}(\bs)\}$ would have made had it been included in the regression.
This means that a fixed effect that has been inadvertently left out will be picked up in the spatial statistical model \eqref{eqn-y-bs}, as a random effect.

\noindent\textbf{Principle 3.2}: \textit{Assume a decomposition of a spatial process into fixed effects plus random effects. While it is not unique, the decomposition must be chosen to conserve variability.}

This principle is a refinement of Principle 2.3 that is adapted here for Spatial Statistics.
The variability of the deterministic-mean component is measured differently from that of a random-error component.
Define the \textit{regional variance} of $\mu(\cdot)$ as,
$$
s^2_\mu=\left(\frac{1}{|D|}\right)\int_D(\mu(\bs)-\overline{\mu})^2\,\text{d}\bs\,,
$$
where $\overline{\mu}=(1/|D|)\int_D\mu(\bs)\,\text{d}\bs$ is the regional average of $\mu(\cdot)$~\citep{lahiri1999}.
Suppose that a stationary covariance function, $C_\delta(\cdot)$, describes the covariability in $\{\delta(\bs):\bs\in D\}$, with $\sigma_\delta^2= C_\delta(\bzero)$; recall that $C_\delta(\bh)=\cov(\delta(\bs+\bh),\delta(\bs))$, for $\bh\in\RR^d$. 
Let $\hat{C}_\delta(\cdot)$ denote the empirical covariance function, and $s_\delta^2=\hat{C}_\delta(\bzero)$.
Then, according to Principle 3.2, \textit{approximately} speaking, $s_\mu^2+s_\delta^2$ should not change as different decompositions given by \eqref{eqn-y-bs} are fitted.
Now suppose that the additional covariate vector, $\bx_{p+1}=(x_{p+1}(\bs_1),\ldots,x_{p+1}(\bs_n))^\top$, is included in the regression, and it is not in the column space of $\bX$.
Then generally, the new empirical covariance function $\hat{C}_\delta(\cdot)$, yields a new estimate $s_\delta^2= \hat{C}_\delta(0)$.
The principle says that this new $s_\delta^2$ should decrease to allow for the additional spatial variability in $\{\mu(\bs):\bs\in D\}$.
(Understandably, the general shape of the new empirical covariance function will change as well.)

How will we ever know which model is better after each is fitted to the same spatial data $\{z(\bs_i)\}$?
Model-selection criteria such as `Akaike Information Criterion,' `Deviance Information Criterion,' and cross-validation will remove bad models but, in the `difficult middle,' there is no way to know whether a graph of $\{z(\bs_i)\}$ versus $\{x_{p+1}(\bs_i)\}$ is showing the behaviour of a component of the mean function or the behaviour of a component of the error process.
Germane to Principle 3.2, it is well known to spatial statisticians that a single simulation of a strongly spatially dependent random effect, $\{\delta(\bs):\bs\in D\}$, can look like deterministic spatial trend.
If a replicate of the data were available, and if the analogous plot showed the same behaviour as seen in the original plot, then $\{x_{p+1}(\bs):\bs\in D\}$ would probably belong in the mean function $\mu(\cdot)$.
From just one realisation, the decomposition \eqref{eqn-y-bs} is not unique, and hence what is one person's mean function could be another person's spatial error.

\subsection{COS is the DNA of Spatial Statistics}

Most of us have heard of, or encountered, \textit{Simpson's Paradox}, which basically says that two variables $x$ and $y$ could be positively correlated but, \textit{conditional} on a third variable $w$, it is perfectly acceptable that $x$ and $y$ be negatively correlated (or uncorrelated, or positively correlated)!
Usually, Simpson's Paradox is expressed in terms of categorical data, which we could do here by defining ordinal categories, or bins, from the ranges of $w,x$, and $y$.

Let a sample $\{(w_i,x_i,y_i):i=1,\ldots,n\}$ be assigned to the pre-defined bins, creating a three-way contingency table with counts in each cell of the table.
A two-way table showing (marginal) dependence between binned $x$ and binned $y$ is obtained by aggregating the counts across the third variable, namely binned $w$.
Then a measure of dependence in the two-way table (e.g., the statistic \textit{gamma}, due to \citealp*{kruskall1954}) could show positive dependence, but the same measure applied to the two-way tables of binned $x$ and binned $y$ \textit{conditional} on each of the values of binned $w$, could all show negative (or no, or positive) dependence.

What can you do about it?
First, understand why it happens, and then spend the rest of your data-science career looking for those lurking variables like $w$ that you may have missed!
It manifests in other settings as well, as I now discuss.

Let $x$ and $y$ be jointly Gaussian random variables and related through the simple linear regression model given by~\eqref{eq:z-i}; the probability model of how $y$ varies with $x$ is as follows: Conditional on $x$, the random variable $y$ is Gaussian with mean and variance, respectively,
\begin{align*}
& \mathrm{E}(y|x)=\mathrm{E}(y)+\{\rho_{xy}\var(y)^{1/2}/\var(x)^{1/2}\}\{x-\mathrm{E}(x)\}\,,\\
& \var(y|x)=\var(y)\{1-\rho_{xy}^2\}\,,
\end{align*}
where $\rho_{xy}=\corr(x,y)$.

Now consider regression of $y$ on both $x$ and $w$, where again joint Gaussianity of random variables $w,x$, and $y$ is assumed.
To make the calculations easier to follow, in the rest of this subsection consider the simple case where $\mathrm{E}(w)=\mathrm{E}(x)=\mathrm{E}(y)=0$, and $\var(w)=\var(x)=\var(y)=1$.
Then the covariances, $\cov(x,y)$, $\cov(w,y)$, and $\cov(x,w)$ are identical to correlations, which are denoted here as $\rho_{xy},\rho_{wy}$, and $\rho_{xw}$, respectively.
The regression lines to be compared are,
\begin{equation}
\label{eq:regression-lines}
\mathrm{E}(y|x,w)=\left\{\frac{\rho_{wy}-\rho_{xy}\rho_{xw}}{1-\rho_{xw}^2}\right\}w+\left\{\frac{\rho_{xy}-\rho_{wy}\rho_{xw}}{1-\rho_{xw}^2}\right\}x\,,
\end{equation}
and
$$
\mathrm{E}(y|x)=0+\rho_{xy}\cdot x\,.
$$
Then, conditional on $w$,
\begin{align}
\label{eq:cond-on-w}
\corr(x,y|w)&=\left\{\frac{\rho_{xy}-\rho_{wy}\rho_{xw}}{1-\rho_{xw}^2}\right\}\frac{\var(x|w)^{1/2}}{\var(y|w)^{1/2}}\nonumber\\
&=\left\{\frac{\rho_{xy}-\rho_{wy}\rho_{xw}}{(1-\rho_{xw}^2)^{1/2}(1-\rho_{wy}^2)^{1/2}}\right\}\,,
\end{align}
which is to be compared to $\corr(x,y)=\rho_{xy}$.

If $w$ has zero correlation with both $x$ and $y$, then from \eqref{eq:cond-on-w}, $\corr(x,y|w)=\corr(x,y)=\rho_{xy}$, which is an intuitively reasonable result.
In general, $\rho_{wy}\neq0$ and $\rho_{xw}\neq0$, so from \eqref{eq:regression-lines} it is clear that a lurking variable $w$ can wreak havoc on any honest attempt to interpret a simple linear regression of $y$ on $x$.
But, what does Simpson's Paradox have to do with spatial statistics?

The answer is `plenty,' if you think about $w$ as being a variable that describes a range of geographical strata.
For example, the Australian Bureau of Statistics divides Australia up into \textit{small areas}, at different levels of aggregation.
A study of $y=$mean weekly income in NSW, Australia, regressed on selected demographic variables $\bx$ at the finest level of aggregation, was given in \cite{burden2015}.
Given the great divide in Australia between `city' and `country,' the most obvious variable $w$ to choose is: $w=1$ if the small area is in Greater Sydney (city), and otherwise $w=0$ (country).
In this set-up, it is easy to imagine that the two regressions, $\mathrm{E}(y|\bx,w=1)$ and $\mathrm{E}(y|\bx, w=0)$, would result in more interpretable results than the single regression, $\mathrm{E}(y|\bx)$.
(In fact, \citealp{burden2015}, used Principle 3.2 and captured the `geography' by using a spatial error term rather than a geographic covariate $w$.)

Simpson's Paradox is potentially everywhere in the spatial context, because regressions of $y$ on $x$ can be done at different levels of spatial aggregation.
A regression of $y$ on $x$ at the finest level of aggregation may show positive dependence, but when the variables are aggregated to a coarser level, the regression may show negative (or no, or positive) dependence!
In Sociology, this phenomenon has been called the \textit{ecological effect}~\citep{robinson1950}, an unfortunate and misleading name that has no direct connection to Ecology.
In Geography, the combination of Simpson's Paradox and the ecological effect has been called the Modifiable Areal Unit Problem, or MAUP; a spatial statistical discussion of MAUP is given in \cite{cressie1996}.

Because it is so natural to aggregate processes over space, meta-data in the spatial data set should include the \textit{spatial support} of each datum.
Define $y(B)=(1/|B|)\int_B y(\bs)\,\text{d}\bs$; then $y(B)$ has \textit{spatial support} $B$ with volume, $|B|=\int_B\text{d}\bs>0$.
In geostatistics, $B$ is called a \textit{block}, and we say that $y(B)$ is an aggregation (or block average) of the original process. 
Then a \textit{change-of-support} (COS) is said to have occurred from point support to spatial support $B$, with a corresponding change of statistical properties.
Data on mutually exclusive supports, $\{B_1,\ldots,B_n\}$, are typically represented as:
$$
z(B_i)= y(B_i)+\varepsilon_i\,;\,i=1,\ldots,n\,.
$$
The simplest model would be $\{\varepsilon_i:i=1,\ldots,n\}$ iid $\textrm{N}(0,\sigma_\varepsilon^2)$, but modifications to account for the protocols of measurement and the possible overlap of the supports $\{B_i\}$ have been developed~\citep[e.g.,][]{wikle2005}.

The two earlier principles of Spatial Statistics (nearby things are more alike; and decompose spatial variability into fixed effects plus random effects) are joined here by a change-of-support principle. 

\noindent\textbf{Principle 3.3}: \textit{The variance of the aggregation, $y(B)$, is a decreasing function of the volume, $|B|$.}

\noindent This is the most important of a number of COS properties~\citep[e.g.,][]{gotway2002}, whose discussion I have left for a future time.

Let $y(\bs)= y_0+\delta(\bs)$, for $\bs\in D$, where $y_0$ is a non-degenerate random variable with possibly non-zero mean and independent of $\{\delta(\bs):\bs\in D\}$.
Then for $B\subset D$,
\begin{equation}
\label{eq:var-yb}
\var(y(B)))=\sigma_0^2+\var(\delta(B))\,,
\end{equation}
where $\sigma_0^2=\var(y_0)>0$.
Therefore, if $\var(\delta(B))$ decreases to $0$ as the volume $|B|$ increases, $\var(y(B))$ decreases to $\sigma_0^2>0$.
It is this type of behaviour that is of interest to geoscientists analysing remote sensing data.
In that literature~\citep[reviewed in][]{cressie2018}, they distinguish between two types of error, `systematic error' and `random error,' as follows.

\noindent\textit{Random error}: An error is a random error if the average of a collection of $n$ of them has variability that decreases to zero like $1/n$, as $n$ becomes large.

\noindent\textit{Systematic error}: An error is a systematic error if the average of a collection of $n$ of them has variability that does not decrease to zero, as $n$ becomes large.

These might be considered `verbal working definitions,' but a statistical scientist looks at these and tries to find an appropriate probability model.
The ones I give below are building blocks for parts of a HM.
In practice, the most difficult aspect is to know which errors are of which type and how to group them together for inference.

For `random error,' the obvious probability model is: Let $\delta(\bs_1),\ldots,\delta(\bs_n)$ be iid random variables with mean $0$ and variance $\sigma_\delta^2$.
Now the average, $\overline{\delta}=(\sum_{i=1}^n \delta(\bs_i))/n$, has $\mathrm{E}(\overline{\delta})=0$ and $\var(\overline{\delta})=\sigma_\delta^2/n$, which decreases to zero like $1/n$.
In a `data-rich' situation and with enough averaging of the data, random errors can be shown to be annihilated by the averaging (using the law of large numbers).
In a spatial setting, Principle 3.1 suggests that the independence assumption between the $\{\delta(\bs_i)\}$ is not appropriate.
Under increasing-domain asymptotics, it can be shown that $\textrm{var}(\overline{\delta})$ still decreases like $1/n$ \citep{cressie1993}, however strong spatial dependence can make the errors look more systematic than random \citep{morris1984}.

For `systematic error,' an obvious probability model is a random-effects model: Let $e(\bs_1),\ldots,e(\bs_n)$ be written as
$$
e(\bs_i)=y_0+\delta(\bs_i)\,;\quad i=1,\ldots,n\,,
$$ 
where $y_0$ has mean 0 and variance $\sigma_0^2>0$, and $y_0$ and $\{\delta(\bs_i)\}$ are independent.
Now, if $\{\delta(\bs_i)\}$ are iid mean $0$ and variance $\sigma_\delta^2$, then the average error $\overline{e}=\sum_{i=1}^n e(s_i)/n$ has variance,
$$
\var(\overline{e})=\sigma_0^2+\sigma_\delta^2/n\,.
$$
This does not decrease to zero as $n$ becomes large, and hence $e(\cdot)$ is a form of systematic error.
\citet{zhang2019} used spatial dependence in the $\{\delta(\bs_i)\}$ and an additive random effect $y_0$ with mean $0$, as part of the OCO-2 calibration of remote sensing data to `ground-truth' data.

Consider spatial prediction errors, $\{\hat{y}(\bs_i)-y(\bs_i): i=1,\ldots,n\}$, which are located at $\{\bs_1,\ldots,\bs_n\}$ in the spatial domain $D$.
Typically, these prediction errors have individual means that are not zero, and it is often the case in Spatial Statistics that there is no replication to estimate them.
A way out of the conundrum is to assume exchangeability \citep[e.g.,][Section 3.17]{spiegelhalter2004}, which results in a spatial statistical model that exhibits both systematic error and random error; see \citet[Rejoinder]{cressie2018}.

\section{Epilogue}

In the previous sections, I presented five principles for Statistical Science and three special ones for Spatial Statistics.
In what follows, I add a ninth principle that speaks to all of Science, not just Data Science, and I give some concluding remarks.

\subsection{You can't add apples and oranges}

There is one further principle that has been my `rock' in the diverse applied-statistics projects in which I have participated.
I once saw a cartoon that showed a small town's `Welcome' sign, and it looked something like this:
\begin{center}
\begin{tabular}{lcrl}
\multicolumn{3}{c}{\textbf{Centerville}}&\\
Population&:&2,390&\\
Elevation&:&862&\!\!feet\\
Total&:&3,252&
\end{tabular}
\end{center}
Keeping track of units is a fundamental part of all of science, including Data Science.

\noindent\textbf{Principle 4.1}: \textit{Only add quantities that have the same units. Multiply quantities that have different units, and cancel units from the product whenever possible.
Take logs, exponents, and other special functions of unit-free quantities.}

The first part of the principle is illustrated with the cartoon I referred to above.
You might say you would never add apples and oranges but, if in \eqref{eq:bz}, $z$ is measured in petagrams (Pg) of carbon in Earth's atmosphere and $x_1=1$, $x_2=t$ (in years), and $x_3=t^2$, then a unit analysis using the second part of the principle would reveal that regression coefficient $\beta_1$ is in units of Pg, $\beta_2$ is in Pg/yr, and $\beta_3$ is in Pg/yr$^2$.
Principle 4.1 is respected, since it is the regression components $\{\beta_k x_k\}$ that are added.
However, it is the regression coefficients $\{\beta_k\}$ that are often interpreted, so I often standardise the covariates by, respectively, subtracting their averages and dividing by their sample standard deviations.
Then, after this standardisation, $\beta_1,\beta_2,\ldots,\beta_k$ all have the same units as those of $z$.

Statistical scientists know that probabilities have no units.
Using Principle 4.1, a unit analysis of the fundamental probability equation, $\int f(y)\,dy=1$, where $f(\cdot)$ is the probability density function of the random variable $y$, whose units are, say, Pg of carbon, reveals that $f(y)$ has units of (Pg)$^{-1}$.
Using the same principle, $\mathrm{E}(y)$ has units of Pg, and so forth.
While probability density functions have units, cumulative distribution functions and probability mass functions do not.

The last part of Principle 4.1 applies to any special function that can be expressed as a Taylor series.
The most common ones in science are logs and exponents, whose Taylor series are
$$
\log_e(1+x)=x-x^2/2+x^3/3-\ldots
$$
and
$$
e^x=1+x+x^2/2!+x^3/3!+\ldots\,,
$$
which only make sense when $x$ has no units; see the first part of Principle 4.1.
Euler's number, $e$, has no units, because $e=\lim_{n\rightarrow\infty}(1+1/n)^n$.

Counts, percentages, and correlations have no units.
Also, the Box-Cox transformation~\citep*{box1964}, 
$$
g_\lambda(x)=
\begin{cases}
(x^\lambda-1)/\lambda\,,&\text{ for }\lambda\in(-\infty,0)\cup(0,\infty)\\
\log x\,,&\text{ for }\lambda=0
\end{cases}\,,
$$ 
only makes sense if $x$ has been rendered unit-free.
One way to accomplish this is to specify $x$ relative to a given standard.

Every term in the process model (and data model) in a HM, should be assigned their rightful units.
The scientific models that make up parts of the process model are sometimes derived theoretically, sometimes empirically.
Beware of a `scientific constant.'
It may be an (estimated) regression coefficient, in which case its units (the ratio of the dependent variable's units to the corresponding covariate's units) are key, since that `constant' might change if the units change.

In conclusion, it is a critical part of every collaboration to do a `unit analysis,' which avoids obvious mistakes made by confusing different parts of different systems of units (e.g., metric and imperial), as well as the more subtle ones discussed above.

\subsection{These are my nine principles}

\noindent\textbf{Principle 2.1}: \textit{Establish a true model (TM), perhaps different from the scientist's working model (WM). Critically, compute the TM-distributional properties of the WM estimators.} [All models are wrong $\ldots$ but some are wrong-er than others.]

\noindent\textbf{Principle 2.2}: \textit{Build statistical models conditionally, through a data model and a process model. Infer the unknown process from the predictive distribution.} [What you see is not what you want to get.]

\noindent\textbf{Principle 2.3}: \textit{In any well defined statistical model, there is conservation of variability.} [Geophysicists conserve energy but what do data scientists conserve?]

\noindent\textbf{Principle 2.4}: \textit{Seek a transformation of the scientific process where all components of variation act and interact additively.} [The holy grail: all scales of variation are additive.]

\noindent\textbf{Principle 2.5}: \textit{When building probability models, look carefully where zero probabilities are assumed (perhaps implicitly) and, with the same care, move appropriate probabilities away from zero.
Calculate joint probabilities from products of conditional (not marginal) probabilities, unless entropy is maximal.} [Could swans be red?]

\noindent\textbf{Principle 3.1} \textit{Everything is related to everything else, but near things are more related than distant things \citep{tobler1970}.} [Patches in close proximity are commonly more alike $\ldots$]

\noindent\textbf{Principle 3.2}: \textit{Assume a decomposition of a spatial process into fixed effects plus random effects. While it is not unique, the decomposition must be chosen to conserve variability.} [What is one person's mean function could be another person's spatial error.]

\noindent\textbf{Principle 3.3}: \textit{The variance of the aggregation, $y(B)$, is a decreasing function of the volume, $|B|$.} [COS is the DNA of Spatial Statistics.]

\noindent\textbf{Principle 4.1}: \textit{Only add quantities that have the same units. Multiply quantities that have different units, and cancel units from the product whenever possible.
Take logs, exponents, and other special functions of unit-free quantities.} [You can't add apples and oranges.]

\subsection{A disclaimer}

These nine principles of Statistical Science are personal, leading to a certain amount of self-referencing, but I hope others find them useful.
There are no theorem-proof results, but there are back-of-the-envelope calculations that I use to justify the principles in simple settings.
At the very least, they should give data scientists boundaries in their analytics that should be respected, and criteria by which supervised and unsupervised machine-learning methods could be assessed.

\subsection{Happy birthday!}

I once introduced Adrian Baddeley at a conference session I was chairing (ASC2010, Fremantle, Western Australia) as a national treasure.
I repeat it here, and I wish him a very happy 65th birthday!

\medskip

\noindent\textbf{Acknowledgements}: This research was supported by the Australian Research Council under Discovery Project DP190100180.
The Sydney Business School, University of Wollongong, generously provided an environment where most of this article was written.
The referees have generously given comments that have helped me clarify the exposition of my nine principles.
My thanks go to Dr B. Maloney for his skillful technical assistance.
I would like to express my appreciation to Dr J. Wong for insightful discussions on this and other material.

%\newpage

 \bibliography{DSbib}
  
\end{document}